# Orientational coupling between the vortex lattice and the crystalline lattice in a weakly pinned $Co_{0.0075}NbSe_2$ single crystal

Somesh Chandra Ganguli, Harkirat Singh, Rini Ganguly, Vivas Bagwe, Arumugam Thamizhavel and Pratap Raychaudhuri[a]

*Department of Condensed Matter Physics and Materials Science, Tata Institute of Fundamental Research, Homi Bhabha Road, Colaba, Mumbai 400005, India.*

We report experimental evidence of strong orientational coupling between the crystal lattice and the vortex lattice in a weakly pinned Co-doped $NbSe_2$ single crystal through direct imaging using low temperature scanning tunneling microscopy/spectroscopy. When the magnetic field is applied along the six-fold symmetric *c*-axis of the $NbSe_2$ crystal, the basis vectors of the vortex lattice are preferentially aligned along the basis vectors of the crystal lattice. The orientational coupling between the vortex lattice and crystal lattice becomes more pronounced as the magnetic field is increased. This orientational coupling enhances the stability of the orientational order of the vortex lattice, which persists even in the disordered state at high fields where dislocations and disclinations have destroyed the topological order. Our results underpin the importance of crystal lattice symmetry on the vortex state phase diagram of weakly pinned Type II superconductors.

[a] pratap@tifr.res.in

# 1. Introduction

The vortex lattice (VL) in a type II superconductor provides a versatile model system to study the interplay of interaction and pinning[1,2,3]. So far, most descriptions of the vortex state in homogeneous 3 dimensional (3D) superconductors involves the vortex-vortex interactions, which stabilize a hexagonal Abrikosov vortex lattice, and random pinning of vortices by crystalline defects, which tend to destroy this order by pinning the vortices at random positions[4,5,6,7]. For weakly pinned Type II superconductors, these theories predict that the topological defect free VL undergoes an order to disorder transition (ODT) through proliferation of topological defects[8] (TD) as one approaches the superconductor-normal metal phase boundary. These TD relax the hexagonal order which make it easier for vortices to accommodate the random pinning potential thereby enhancing the effective pinning. Thus the ODT manifests as a sharp increase in the critical current or a decrease in the real part of ac susceptibility ($\chi'$), the so called "peak effect" [9], which has been widely studied in weakly pinned superconducting crystals[1,10].

In principle, the VL can also couple with the symmetry of an underlying substrate. In superconductors with artificially engineered periodic pinning, this coupling has been shown to give rise to interesting matching effects, where the VL gets oriented in specific direction with respect to the pinning potential when the lattice constant is commensurate with the pinning potential[11]. In single crystals of conventional superconductors, barring one exception[12], most theories dealing with the vortex phase diagram[4,5,6,7] consider the VL to be decoupled from the crystal lattice (CL) except for the random pinning potential created by defects which hinders the relative motion between two. However, in cubic and tetragonal systems, it has been theoretically[13] and experimentally[14,15] shown that non-local corrections to the vortex-vortex interaction can carry the imprint of crystal symmetry. Recent neutron diffraction experiments[16] on Nb single crystal also show that the structure of the VL varies depending on the symmetry of the crystalline axis along which the magnetic field is applied. Therefore the influence of the symmetry of the CL on the VL and consequently its effect on the ODT needs to be explored further.

In this paper we investigate the coupling between the symmetry of the VL and CL in $Co_{0.0075}NbSe_2$, a Type II superconductor[17] with hexagonal crystal structure. The intercalated Co atoms act as pinning centers, thus providing us a control on the strength of pinning[18]. Our crystal is in the weak-pinning limit, which we functionally define as the pinning range where a topological defect free hexagonal ground state of the VL is realized at low temperatures and low fields. By simultaneously imaging the VL and CL using a scanning tunneling microscope (STM) operating down to 350 mK we investigate the orientation of the VL with respect to the CL. The central result of this paper is that while the VL is always preferentially oriented along the unit cell vector in the layer when the magnetic field is applied along the six-fold symmetric *c*-axis of the crystal, at low fields, the VL lattice can get locked into a metastable state where large randomly oriented domains can be realized. The orientational coupling of the VL with the crystal lattice strongly affects the disordering of the VL and provides a backbone for the robust orientational order of the VL across the ODT.

## 2. Experimental Methods

The single crystal with nominal composition $Co_{0.0075}NbSe_2$ was grown by iodine vapour transport method starting with stoichiometric amounts of pure Nb, Se and Co, together with iodine as the transport agent. Details of sample preparation have been published elsewhere[22]. Energy dispersive X-ray (EDX) analysis was performed on the single crystal to confirm the presence of Co atoms. The EDX spectra showed a Co concentration slightly smaller than the nominal one, varying between 0.5-0.55 atomic %, on different points of the single crystal. However, since precise determination of the composition below 1 at. % is limited by the resolution of EDX we refer to the nominal composition in the rest of the paper. The VL and the CL were imaged using a home-built low temperature STM[19] fitted with a superconducting solenoid operating down to 350 mK. Prior to STM measurements the crystal is cleaved *in-situ* at room temperature, in a vacuum better than $1\times10^{-7}$ mbar. $NbSe_2$ has a layered hexagonal crystal structure with each unit cell consisting of two sandwiches of hexagonal Se-Nb-Se layers. Thus the crystal cleaves between the weakly coupled neighboring Se layers exposing the hexagonal Se terminated surface. STM topographic images were captured at various locations to identify an atomically smooth surface. Fig. 1(a) shows one such region (with surface height variation < 2 Å) over an area of 1.5 μm × 1.5 μm close to the center of the crystal. Atomic resolution images were subsequently captured at various points within this area to determine the unique orientation of the in-plane crystallographic axes (Fig. 1(a) (i)-(iii)). The 3 × 3 charge density wave (CDW) modulation is also visible in the atomic resolution images, though it is blurred due to the presence of Co dopant atoms. The Fourier transform of the atomic resolution image (Fig. 1(a)-(iv)) shows 6 symmetric sharp Bragg spots corresponding to the Se lattice and 6 diffuse spots corresponding to the CDW. All VL images other than those shown in Fig. 2 (b) and (c) were performed within the area shown in Fig. 1(a).

To image the VL, the differential tunneling conductance, $G(V) = dI/dV$, between the tip and the sample is measured as the tip scans the surface at a fixed bias voltage (in constant current mode), $V \sim 1.2$ mV, with a d.c. tunneling current of 50 pA. $G(V)$ is measured by adding a small a.c. modulation voltage ($V_{mod}$) with frequency 2 kHz and amplitude 100 µV to the bias voltage and detecting the resulting modulation in the tunneling current ($I_{mod}$) using a lock-in amplifier, such that, $G(V) \approx I_{mod}/V_{mod}$. The left panel of Fig. 1(b) shows the spatial variation of conductance acquired at 1 kOe and at 350 mK. To improve the contrast, the conductance maps are plotted as a function of $\Delta G(V)$, which is obtained by subtracting from $G(V)$ a constant background value equal to the minimum conductance over the imaged area. The conductance map shows well defined minima (dark spots) corresponding to the position of the vortex cores. The right panel of Fig. 1(b) show $G(V)$ as a function of $V$ at specific locations by sweeping the bias voltage after switching off of STM feedback loop. To avoid a voltage jump at the beginning of the voltage sweep and to increase the resolution, before switching of the feedback loop, the bias voltage is set to the highest value of the sweep voltage (3 mV) and the d.c. tunneling current is increased to 120 pA. The $G(V)$-$V$ curves away from the vortex core consist of well resolved coherence peaks close to the superconducting energy gap and a minimum at zero bias characteristic of a superconductor, whereas inside the normal core they are flat and featureless. Since while acquiring the VL images the bias voltage ($V \sim 1.2$ mV) is kept close to the coherence peak, each vortex manifest as local minimum in the conductance map. Fig. 1(c) shows a typical VL imaged at 10 kOe over the entire area shown in Fig 1(a). To identify topological defects in the VL, we Delaunay triangulate[22] the VL and determine the nearest neighbor coordination for each flux lines. Topological defects typically manifest as points with 5-fold or 7-fold coordination. Complementary measurement of the bulk pinning properties were performed from a.c susceptibility measurements in a home built a.c.-susceptometer fitted with a superconducting solenoid. The a.c field drive was fixed at 60 kHz with an amplitude of 10 mOe. It has earlier been established that for this amplitude the response is in the linear regime.

## 3. Results and Discussion

We first focus on the VL created at a relatively low field of 2.5 kOe in the zero field cooled state (ZFC) where the magnetic field is applied after cooling the sample to the lowest temperature. Fig. 2(a)-(c) show representative images of the ZFC VL at 350 mK imaged over 1.5 µm × 1.5 µm on three different areas of the surface. While Fig. 2(a) was imaged on the same area as that shown in Fig. 1(a), Fig. 2(b) and (c) were performed on two other similar atomically smooth areas on the crystal. We observe that the VL is oriented along different directions at different locations with no apparent relation with the orientation of the crystalline lattice. In addition, in Fig. 1(c) we observe a domain boundary created by a line of dislocations (pairs of adjacent lattice points with 5-fold and 7-fold coordination), with the VL having different orientation on two sides of the boundary. These observations are consistent with earlier Bitter decoration experiments[20] where large area images of the ZFC VL revealed large randomly oriented domains, albeit at much lower fields.

We now show that such randomly oriented domains do not represent the equilibrium state of the VL. It has been shown that shaking the VL through a small magnetic perturbation, forces the system out of metastable states causing a dynamic transition to its equilibrium configuration[21,22,23]. In our case, we did not observe any change when the ZFC VL is perturbed at 350 mK with a magnetic field pulse of 300 Oe by ramping up the field to 2.8 kOe over 8 sec followed by a dwell time of 10 sec and ramping down over 8 sec to its original value[24]. However, when the pulse is applied after heating the sample to a temperature higher than 1.5 K the VL gets oriented along the orientation of the crystal lattice (Fig. 2(d)-(f)). For the VL in Fig. 2(f), in addition, this process annihilates the domain boundary. At 350 mK we need to cycle the VL up to a much higher field in order to orient the domains along the CL. In Fig. 3, we show the VL at the same position as the magnetic field is increased from 2.5 kOe to 7.5 kOe and then decreased to 2.5 kOe. We observe that the VL orientation gradually rotates with increasing field and at 7.5 kOe the VL is completely oriented along the CL. Upon decreasing the field, the VL maintains its orientation and remains oriented along the CL. We also observed that the orientation of the VL does not change when the crystal is heated up to 4.5 K without applying any magnetic field perturbation. Therefore, the domain structure in the ZFC VL corresponds to a metastable state, where different parts of the VL get locked in different orientations.

To explore if this orientational ordering leaves its signature on the bulk pinning properties of the VL we performed ac susceptibility measurements on the same crystal (Fig. 2(g)) using three protocols: In the first two protocols, the VL is prepared in the field cooled (FC) and ZFC state respectively (at 2.5 kOe) and the real part of susceptibility ($\chi'$) is measured while increasing the temperature; in the third protocol the vortex lattice is prepared at the lowest temperature in the ZFC state and the $\chi'$–T is measured while a magnetic field pulse of 300 Oe is applied at regular intervals of 100 mK while warming up. As expected the disordered FC state has a stronger diamagnetic shielding response than the ZFC state representing stronger bulk pinning. For the pulsed-ZFC state, $\chi'$–T gradually diverges from the ZFC warmed up state and shows a weaker diamagnetic shielding response and exhibits a more pronounced dip at the peak effect. Both these show that the pulsed-ZFC state is more ordered than the ZFC warmed up state, consistent with the annihilation of domain walls with magnetic field pulse.

We now investigate the impact of this orientational coupling on the ODT of the VL. As the field is increased further (Fig. 4) the VL remains topologically ordered up to 24 kOe. At 26 kOe dislocations proliferate in the VL, in the form of neighboring sites with 5-fold and 7-fold coordination. At 28 kOe, we observe that the disclinations proliferate into the lattice. However, the corresponding FT show a six-fold symmetry all through the sequence of disordering of the VL. Comparing the orientation of the principal reciprocal lattice vectors with the corresponding ones from the FT of crystal lattice, we observe that the VL is always oriented along the crystal lattice direction. In contrast to our previous studies[22], here an isotropic amorphous vortex glass phase is not realized even after the disclinations have proliferated in the system. This difference reflects the weaker defect pinning in the present crystal[17], which enhances the effect of orientational coupling in maintaining the orientation of the VL along the crystal lattice.

The pertinent question arising from our experiments is, what is the origin of this orientational coupling? Conventional pinning cannot explain these observations, since it requires a modulation of the superconducting order parameter over a length scale of the order of the size of the vortex core, which is an order of magnitude larger that the interatomic separation and the CDW modulation in $NbSe_2$. The likely origin of orientational coupling is from anisotropic vortex cores whose orientation is locked along a specific direction of the crystal lattice. In unconventional superconductors (e.g. $(La,Sr)CuO_4$, $CeCoIn_5$)[25,26] such anisotropic cores could naturally arise from the symmetry of the gap function, which has nodes along specific directions. However even in an s-wave superconductor such as $YNi_2B_2C$ vortex cores with 4-fold anisotropy has been observed, and has been attributed to the anisotropy in the superconducting energy gap resulting from Fermi surface anisotropy[27]. As the magnetic field is increased, the vortices come closer to each other and start feeling the shape of neighboring vortices. The interaction between the vortices in such a situation would also be anisotropic, possessing the same symmetry as that of the vortex core. Thus the interaction energy would get minimized for a specific orientation of the VL with respect to the CL.

To explore this possibility we performed high resolution imaging of a single vortex core at 350 mK. To minimize the influence of neighboring vortices we first created a ZFC vortex lattice at 350 mK (Fig. 5(a)) in a field of 700 Oe, for which the inter vortex separation (177 nm) is much larger than the coherence length. As expected at this low field the VL is not aligned with the CL (Fig. 5(e)). We then chose a square area enclosing a single vortex and measured the full $G(V)$-$V$ curve from 3mV to -3mV at every point on a 64 × 64 grid. In Fig 5(b)-(d) we plot the normalized conductance $G(V)/G(\,3\,mV\,)$ at 3 bias voltages. The normalized conductance images reveal a hexagonal star shape pattern consistent with previous measurements[28,29] in undoped $NbSe_2$ single crystals. Atomic resolution images captured within the same area (Fig. 5(e)) reveals that the arms of the star shape in oriented along the principal directions of the CL. We observe that the star shape is not specifically oriented along any of the principal directions of the VL, which rules out the possibility that the shape arises from the interaction of supercurrents surrounding each vortex core. We believe that the reduced contrast in our images compared to refs. 28,29 is due to the presence of Co impurities which act as electronic scattering centers and smear the gap anisotropy through intra and inter-band scattering.

The six-fold symmetric vortex core structure explains why the ZFC VL gets oriented when we cycle through larger fields. As the magnetic field is increased the vortices come closer and feel the star shape of neighboring vortices. Since the star shape has specific orientation with respect to the CL, the VL also orients in a specific direction with respect to the CL. When the field is reduced, the vortices no longer feel the shape of neighboring vortices, but the lattice retains its orientation since there is no force to rotate it back. Since the 6-fold symmetry of the vortex core here is the same as the hexagonal Abrikosov lattice expected when the vortex cores are circular, we do not observe any field induced structural phase transition of the VL as observed in superconductors where the vortex core has four-fold symmetry[15].

## 4. Conclusion

In conclusion, we have used direct imaging of the crystal lattice and the VL using STM/S to show that the orientation of the VL in a conventional s-wave superconductor is strongly pinned to the crystal lattice. This orientational coupling influences both the equilibrium state at low fields and the order-disorder transition at high field. While at low fields, locally misoriented domains can indeed be observed in the ZFC state, these domains get oriented along the CL when the system is cycled through a larger field. In addition, the persistence of orientational order in the VL at high fields even after proliferation of topological defects, clearly suggests that this coupling can be energetically comparable to the random pinning potential and cannot be ignored in realistic models of the VL in weakly pinned Type II superconductors. We hope that future theoretical studies will quantitatively explore the magnitude of the energy scale of this coupling vis-à-vis vortex-vortex interactions and random pinning and its effect on the vortex phase diagram of Type II superconductors.

*Acknowledgements:* We thank Shobo Bhattacharya for encouragement and support, Victoria Bekeris, Gautam Menon and Herman Suderow for valuable discussions. This work was supported by Department of Atomic Energy, Government of India and Science and Engineering Research Board, Department of Science and Technology through Grant No: EMR/2015/000083.

*Figure captions*

**Figure 1.** (a) The central panel shows topographic image of an atomically smooth surface on the $Co_{0.0075}NbSe_2$ single crystal. The surrounding panels (i)-(iii) show representative atomic resolution images on three different area within this region; the direction of the basis vectors in the hexagonal crystal lattice plane comprising of selenium atoms are shown with green arrows. Panel (iv) shows the Fourier transform corresponding to (i); the Bragg spots corresponding to the atomic lattice and CDW in the FT and shown by arrows. We use the convention (length)$^{-1}$ for the scale-bar on the FT. The topographic images are acquired in constant current mode, with tunneling current of 150 pA and voltage bias of 10 mV. (b) (left) Differential conductance map measured at a bias voltage of 1.2 mV showing the ZFC VL image at 350 mK and 1 kOe; the dark spots correspond to the location of the vortex cores. (right) $G(V)$ vs $V$ corresponding to locations A, B and C marked on the left panel. (c) Differential conductance map showing the vortex lattice at 10 kOe and 350 mK.

**Figure 2.** (a)-(c) Differential conductance maps showing the ZFC VL images (1.5 μm × 1.5 μm) recorded at 350 mK, 2.5 kOe at three different places on the crystal surface. (d)-(f) VL images at the same places as (a)-(c) respectively after heating the crystal to 1.5 K (for (d)) or 2 K ((e) and (f)) and applying a magnetic pulse of 300 Oe. Solid lines joining the vortices show the Delaunay triangulation of the VL and sites with 5-fold and 7-fold coordination are shown as red and white dots respectively. The direction of the basis vectors of the VL are shown by yellow arrows. In figure (c) a line of dislocations separate the VL into two domains with different orientations. The right inset in (d)-(f) show the orientation of the lattice, imaged within the area where the VL is imaged. (g) Susceptibility ($\chi'$) as a function of temperature (T) measured at 2.5 kOe while warming up the sample from the lowest temperature. The three curves correspond to $\chi'$-T measured after

field cooling the sample (FC-W), after zero field cooling the sample (ZFC-W) and zero field cooling the sample and then applying a magnetic pulse of 300 Oe at temperature intervals of 0.1 K while warming up (pulsed ZFC-W). The y-axis is normalized to the FC-W $\chi'$ at 1.9 K. The measurements are done at 60 kHz using an a.c. excitation field of 10 mOe.

**Figure 3.** Differential conductance maps showing the VL at 350 mK at the same location on the sample surface imaged while cycling the magnetic field. The upper panels show the VL as the field in increased from 2.5 kOe to 7.5 kOe after preparing the VL in the ZFC state at 2.5 kOe. The two lower panels show the VL as the field is decreased from 7.5 kOe. Solid lines joining the vortices show the Delaunay triangulation of the VL. The yellow arrows show the direction of the basis vectors of the hexagonal VL. The inset in the lower right of the VL at 7.5 kOe shows representative topographic image of the atomic lattice within this area. The green arrows show the basis vector of the CL. All VL images were acquired over 1 μm × 1 μm.

**Figure 4.** (a) Representative image of the CL (left) along with its FT (right); the directions of the principal lattice vectors and reciprocal lattice vectors are shown in their respective panels with green arrows. We use the convention (length)$^{-1}$ for the scale-bar on the FT. The CL is imaged is imaged within the area where the VL is imaged. (b)-(d) Conductance maps showing the VL lattice at 22, 26 and 28 kOe (left) along with their FT (right); the middle panels show Delaunay triangulation of the VL where the vertex of each triangle represent the position of the vortex core; sites with 5-fold and 7-fold coordination are shown as red and blue dots respectively. The disclinations (unpaired 5-fold or 7-fold coordination sites) observed at 28 kOe are highlighted with green and purple circles. The directions of the principal reciprocal lattice vectors are shown with yellow arrows.

**Figure 5.** (a) Differential conductance map showing the ZFC VL at 700 Oe and 350 mK. (b) High resolution image of the single vortex (114 nm ×114 nm) highlighted in the blue box in panel (a) obtained from the normalized conductance maps ( G(V)/G(V = 3mV) ) at 3 different bias voltages. The vortex core shows a diffuse star shaped patters; the green arrows point towards the arms of the star shape from the center of the vortex core. (c) Atomic resolution topographic image of the CL imaged within the box shown in (b); the green arrows show the principal directions of the crystal lattice.

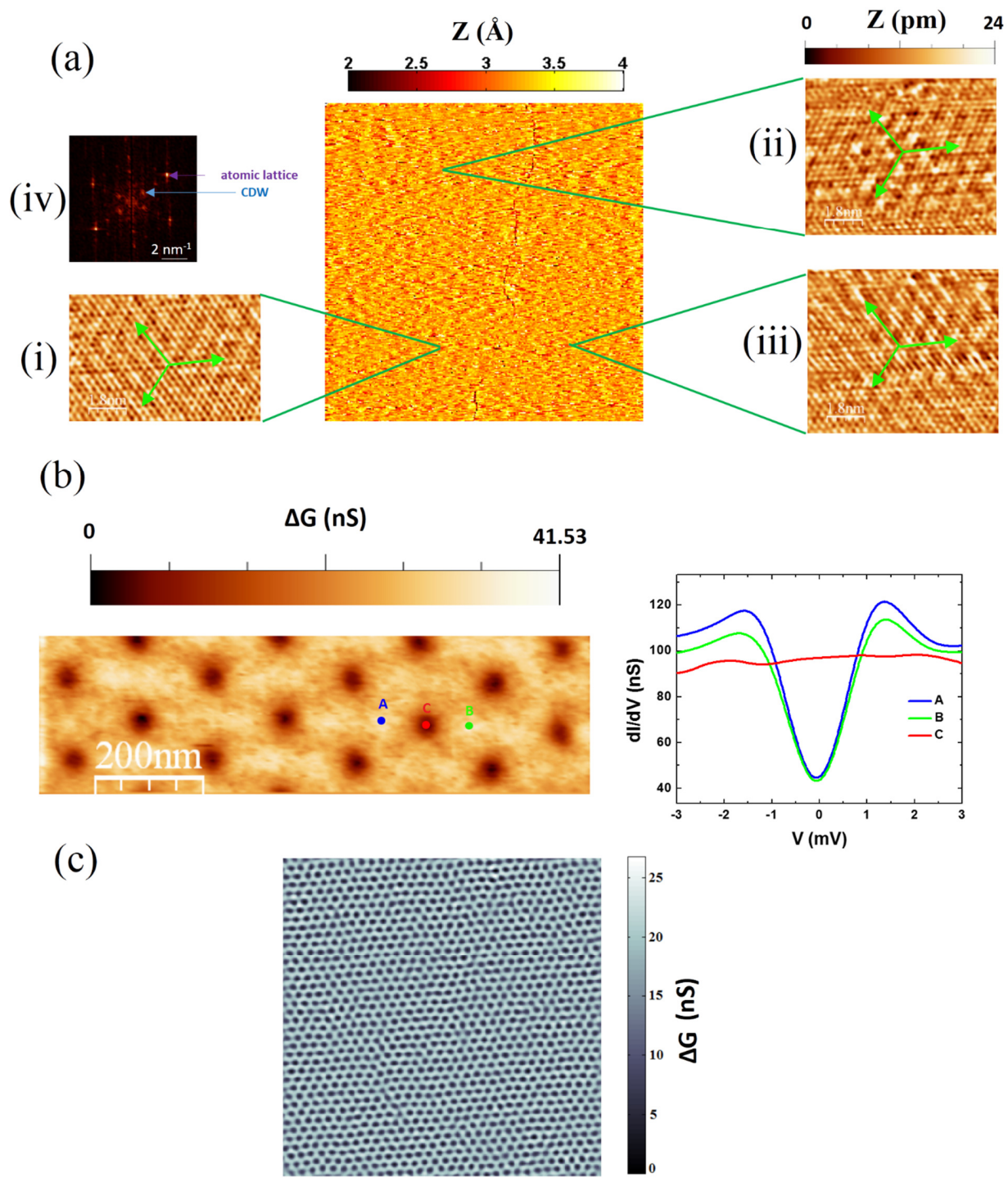

**Figure 1**

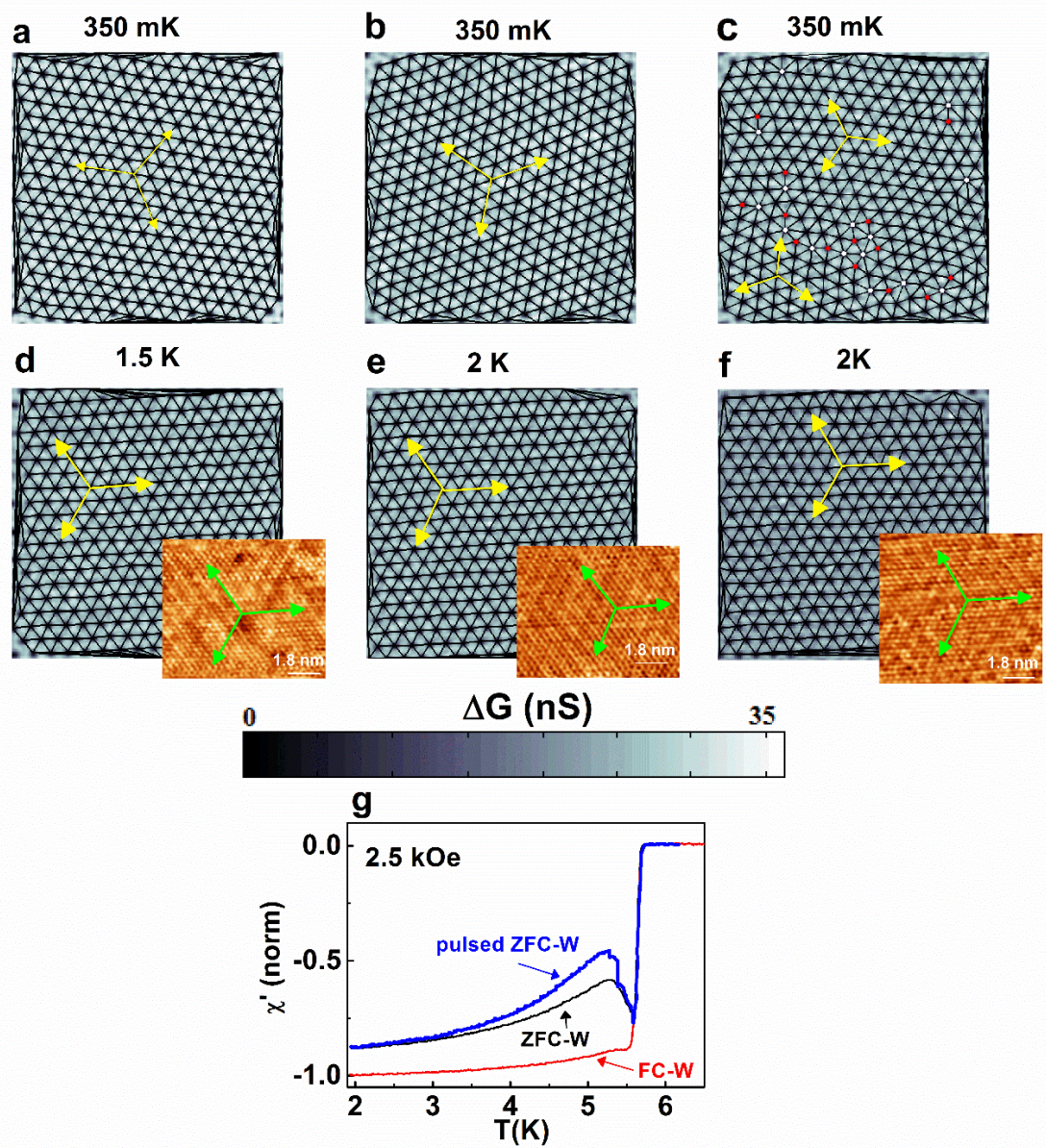

**Figure 2**

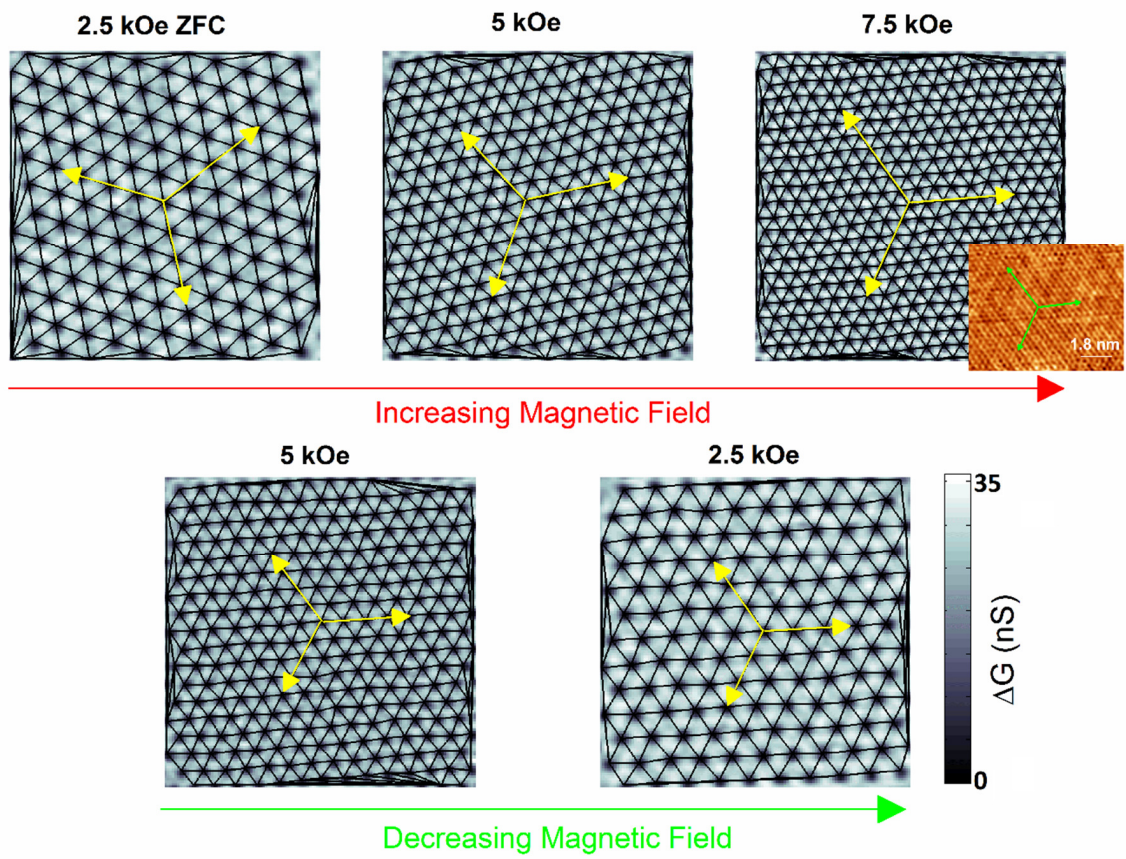

**Figure 3**

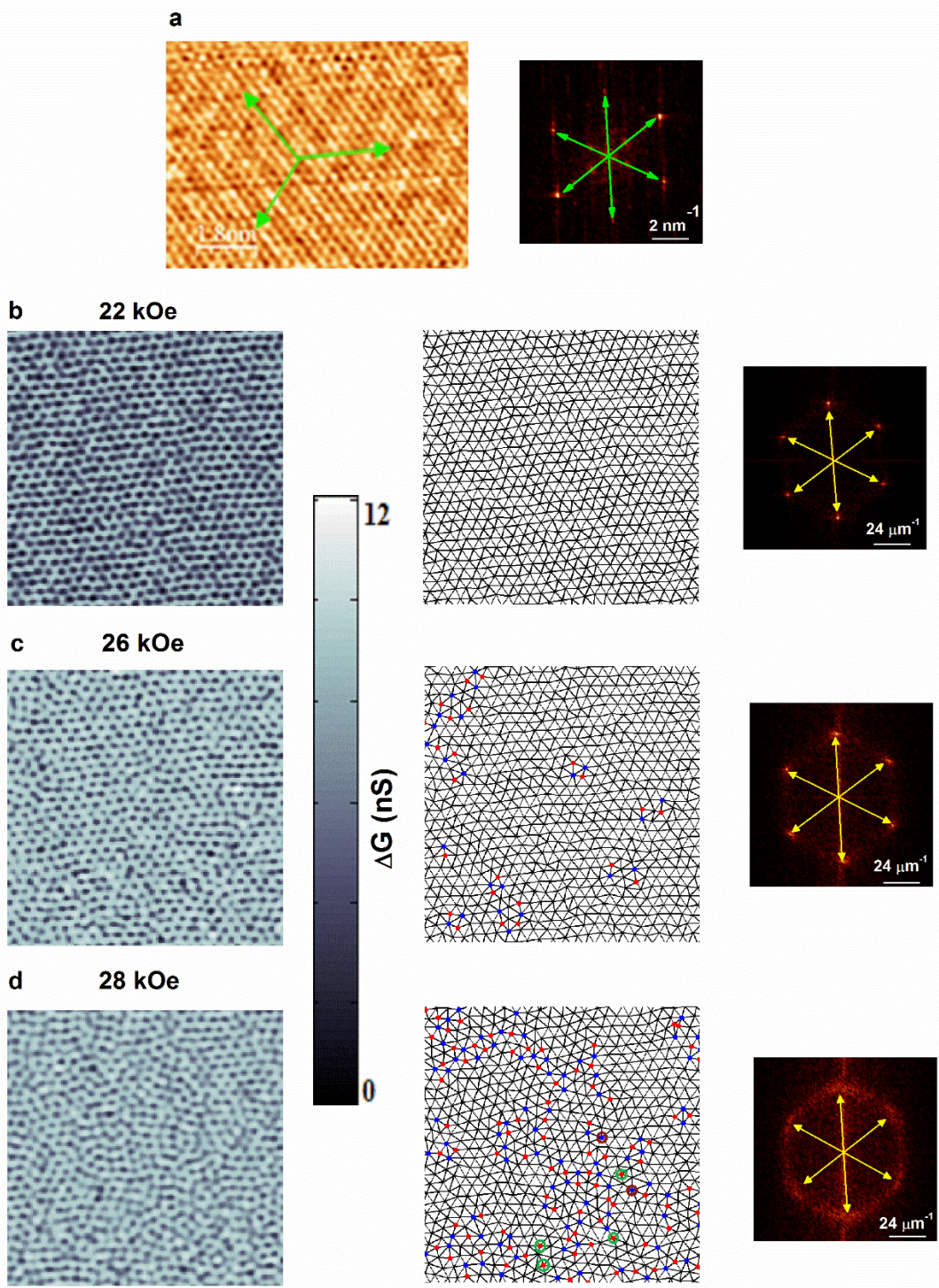

**Figure 4**

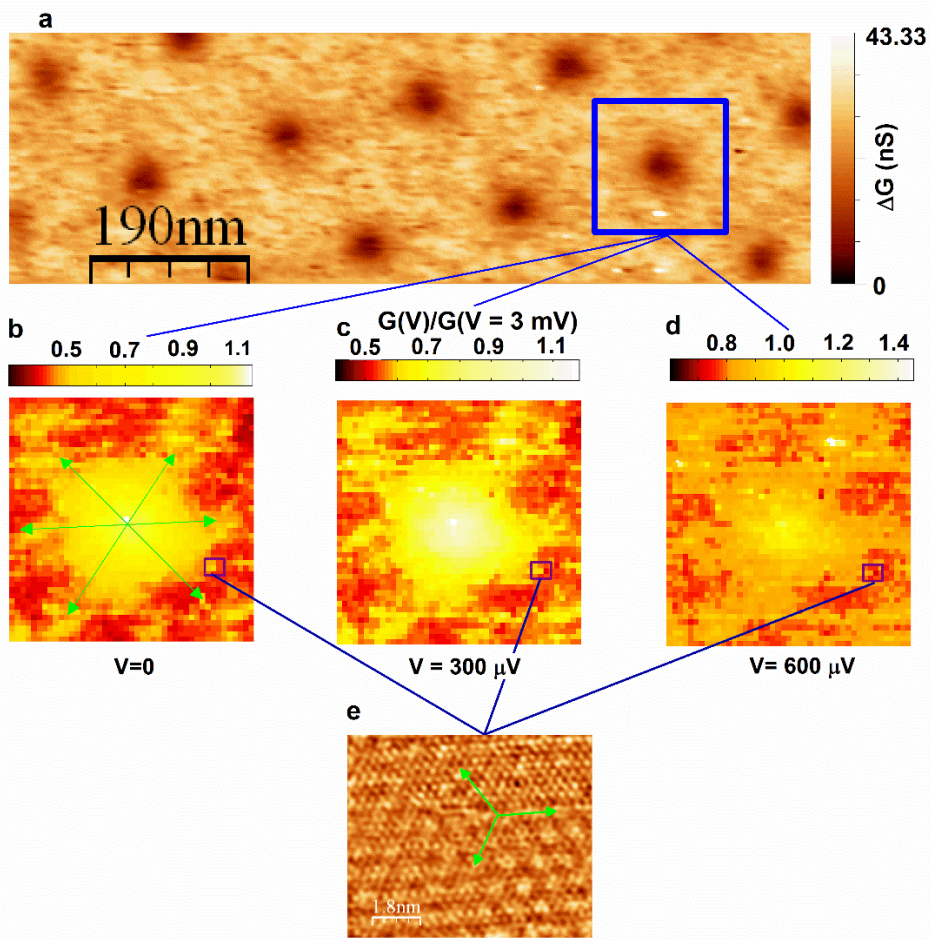

**Figure 5**

# Orientational coupling between the vortex lattice and the crystalline lattice in a weakly pinned $Co_{0.0075}NbSe_2$ single crystal

Somesh Chandra Ganguli, Harkirat Singh, Rini Ganguly, Vivas Bagwe, Arumugam Thamizhavel and Pratap Raychaudhuri[2]

*Department of Condensed Matter Physics and Materials Science, Tata Institute of Fundamental Research, Homi Bhabha Road, Colaba, Mumbai 400005, India.*

## A. Bulk characterization of $Co_{0.0075}NbSe_2$ crystal using a.c. susceptibility

The superconducting $Co_{0.0075}NbSe_2$ crystal was characterized using a home built a.c. susceptometer operating at 60 kHz. The amplitude of the a.c. signal was fixed at 10 mOe. The linearity of the response was previously verified (Ref. 20) for this amplitude.

Fig. 1S shows the real and imaginary part ($\chi'$ and $\chi''$) of a.c. susceptibility in zero field. The crystal shows a sharp superconducting transition with $T_c \sim 5.9$ K. The transition temperature is higher than the sample with the same nominal composition studied in ref. 20, showing that the crystal is less disordered.

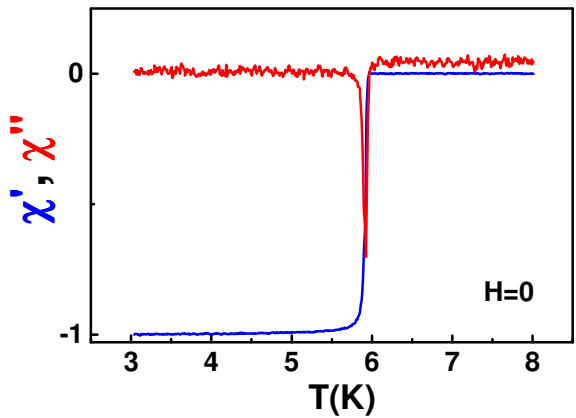

**Figure 1S.** Real ($\chi'$) and imaginary ($\chi''$) part of a.c. susceptibility of the $Co_{0.0075}NbSe_2$ crystal in zero field. Both curves are normalized to the $\chi'$ value at 3 K.

To characterize the bulk VL response $\chi'$ was measured as a function of magnetic field (H) at 1.9 K after cooling the crystal in zero magnetic field. Fig. 2S shows $\chi'$-H for increasing and decreasing field. We observe a pronounced peak effect associated with the ODT, with the $\chi'$ reaching a minimum at $H_p \sim 25$ kOe. The increasing and decreasing curves overlap with each other except close to the peak effect as reported previously in ref. 20. The χ'-T measured for the ZFC state prepared at 1.9 K (in a field of 5 kOe) also shows a sharp peak effect close to the superconducting transition temperature. Consistent with the higher $T_c$ the peak effect is also sharper compared to ref. 20 showing that the crystal is cleaner and hence more weakly pinned.

---

[2] pratap@tifr.res.in

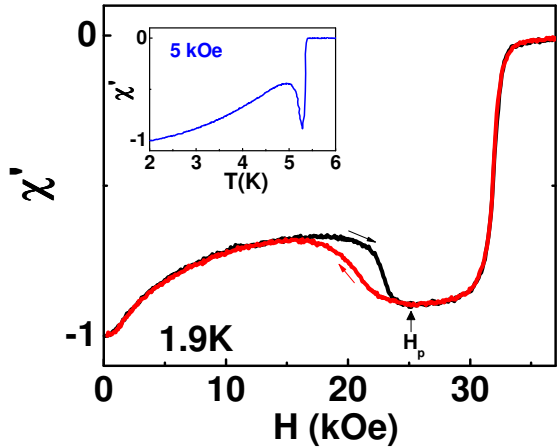

**Figure 2S.** Magnetic field variation of the real part of a.c. susceptibility (χ') normalized to the zero field value, at 1.9 K. The black and red curves correspond to increasing and decreasing fields respectively. (*inset*) Temperature variation of χ' at 5 kOe.

**B. Penetration of the magnetic field pulse into the sample at 350 mK**

Since we do not observe a reorientation of the ZFC vortex lattice at 2.5 kOe at 350 mK when a magnetic field pulse of 300 Oe is applied, it is important to ensure that the applied magnetic field pulse indeed penetrates the sample. To verify this we have imaged the ZFC VL at the same place first at 350 mK first in a field of 2.5 kOe and then after ramping up the field to 2.8 kOe (Fig. 3S). The average lattice constant from both images is calculated from the nearest neighbor bond length distribution of the Delaunay triangulated lattice. Fitting the histogram of the bond lengths to Gaussian distribution, we observe that the average lattice constant decreases by 5% after the magnetic field is ramped up by 300 Oe. This is very close to the value of 5.5% expected if the field fully penetrates in the superconductor, showing that our magnetic field pulse indeed penetrates the sample.

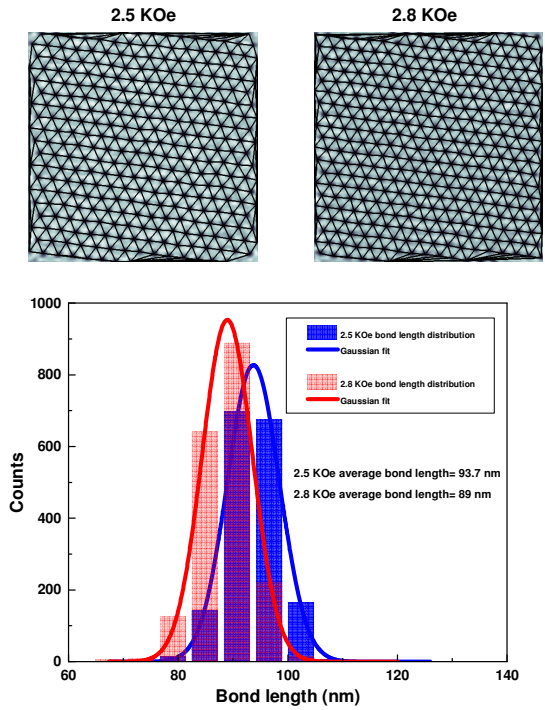

**Figure 3S.** (Top) VL image at 2.5 kOe (left) and 2.8 kOe (right). The solid lines show the Delaunay triangulation of the VL. (Bottom) Histograms showing the nearest neighbor bond length distributions at 2.5 and 2.8 kOe. The solid lines are the Gaussian fits. Spurious bonds at the edge of the images arising from the Delaunay triangulation procedure are ignored in this analysis.